\documentclass{epl}
\title{Scaling and noise-crosscorrelations in a Burgers-like model
for Magnetohydrodynamics}
\shorttitle{Noise-crosscorrelations and Magnetohydrodynamics}
\author{Abhik Basu\inst{1,2}\thanks{E.mail: 
\email{abhik@physics.iisc.ernet.in}}}
\institute{\inst{1} Centre for Condensed Matter Theory, Department of 
Physics, Indian Institute of Science, Bangalore 560012, India,\\
\inst{2} Poorna Prajna Institute of Scientific Research, Bangalore, India.}
\begin{document}
\maketitle
\begin{abstract}
We study the effects of crosscorrelations of noises on the scaling properties
of the correlation functions in a reduced model for Magnetohydrodynamic (MHD) 
turbulence
[A. Basu, J.K Bhattacharjee and S. Ramaswamy [{\em Eur. Phys. J B} {\bf 9},
725 (1999)]. 
We show that in {\em dimension d} 
crosscorrelations with sufficient long wavelength singularity become 
relevant and take the system to the long
range noise fixed point. The crosscorrelations also
affect the ratio of energies of the magnetic and velocity fields
($E_b/E_v$) in the strong coupling phase. 
In dimension $d=1$ the fluctuation-dissipation theorem (FDT)
does not hold in presence of short range crosscorrelations.
We discuss the possible effects of
crosscorrelations on the scaling properties of fully developed MHD turbulence.
\end{abstract}

Numerical studies of MHD turbulence in steady state demonstrate the 
existence of scaling
and  multiscaling properties of the structure functions \cite{abprl,thesis}
which are different from fluid turbulence. The ratio $E_b/E_v\equiv
{\int_{\bf k}\langle \bf b(k).b(-k)\rangle\over\int_k 
\langle v(k).v(-k)\rangle}$ is believed
to be an important parameter in characterising statistical properties of
MHD turbulence: $E_b/E_v=0\Rightarrow$ fluid limit, $E_b/E_v<<1\Rightarrow$
the kinetic regime, $E_b/E_v\sim 1\Rightarrow$ the equipartition regime,
and $E_b/E_v>>1\Rightarrow$ the magnetic regime (dominated by a mean magnetic
field) \cite{mont}. As $E_b/E_v$ increases from 0 to 1, one should be able
to observe a crossover from fluid-like behaviour ro MHD-like behaviour
- this has been confirmed in a recent shell model study \cite{thesis,abrp}.
In a model for MHD
turbulence $E_b/E_v$ appears as a derived quantity depending upon the 
viscosities and the external forcings. So for a theoretical understanding,
it is important to know which parameters influence $E_b/E_v$. 
MHD turbulence is governed by the equations
of magnetohydrodynamics (3dMHD) \cite{mont}. However, for simplicity
we work with the one-dimensional ($1d$) reduced model of MHD \cite{jkb} and
its $d$-dimensional generalisation.
In this letter we analytically examine the scaling properties of the
model equations, driven by stochastic noises. We, in
particular, study the effects of noise-crosscorrelations. 
Our main results, obtained by applying
renormalisation group and self-consistent mode coupling methods on the model
equations (see below) include the existence of a roughening 
transition for $d>2$
for short range noises, the dependence of 
$E_b/E_v$ on the crosscorrelations in the rough phase, and breakdown of the
FDT in $1d$.
The model equations can be easily extended to $d$-dimension by considering
${\partial\over\partial x}\rightarrow \nabla$ and $u,\,b$ as vectors (see
Ref.\cite{decay} for a study of decaying MHD turbulence using the 
$d$-dimensional version of the model of Ref.\cite{jkb}):
\begin{equation}
{\partial {\bf u}\over \partial t}+{\lambda_1\over 2}{\nabla}u^2
+{\lambda_2\over 2} {\nabla}b^2= \nu{\nabla^2}{\bf u} +{\bf f},
\label{jkb1u}
\end{equation}
\begin{equation}
{\partial {\bf b}\over \partial t}+\lambda_3{\nabla}({\bf u.b})=
\mu{\nabla^2\bf b} +{\bf g},
\label{jkb1b}
\end{equation}
where $\nu$ and $\mu$ are fluid and magnetic viscosities respectively,
$\bf f$ and $\bf g$ are external forcing functions.
The Galilean invariance of the Eqs.(\ref{jkb1u}) and (\ref{jkb1b}) requires
that $\lambda_1=\lambda_3$; $\lambda_2$ can be left arbitrary \cite{jkb}.
The external forces $\bf f,g$ are taken to be stochastic, with zero
mean and Gaussian distributions. We choose,
\begin{eqnarray}
\langle f_i({\bf k},t)f_j({\bf -k},0)\rangle&=&2k_ik_jD_1(k)\delta(t),\\
\label{d1eq}
\langle g_i({\bf k},t)g_j({\bf -k},0)\rangle&=&2k_ik_jD_2(k)\delta(t),\\
\label{d2eq}
\langle f_i({\bf k},t)g_j({\bf -k},0)\rangle&=&2D_{ij}({\bf k})\delta(t).
\label{d3eq}
\end{eqnarray}
Care must be taken while fixing the structure of $D_{ij}({\bf k})$: It should
be imaginary and odd in $k$ \cite{jkb} due to the specific parity properties
of $\bf u$ and $\bf b$. We break up $D_{ij}$ into its symmetric and 
antisymmetric parts (it is easy to see that in $1d$ only the symmetric part 
survives): $D_{ij}=D_{ij}^s+D_{ij}^a$ with $D_{ij}^s=
ik_ik_j\tilde{D}({\bf k})$ with $\tilde{D}({\bf k})=-\tilde{D}({\bf -k})$
is the symmetric part and we set the anitisymmetric part $D_{ij}^a=0$ for time 
being. Then the Eqs.\ref{jkb1u} and \ref{jkb1b} can be converted to those of 
Erta\c{s} and Kardar, which describe the properties of drifting
polymers in a medium by the simple transformations ${\bf u}= 
\nabla h,\,{\bf b}=\nabla \phi$ and ${\bf f_u}=\nabla \eta_1,\,{\bf f_b}
=\nabla \eta_2$  
\begin{equation}
{\partial h\over\partial t}+{\lambda_1\over 2}(\nabla h)^2
+{\lambda_2\over 2}(\nabla \phi)^2=\nu\nabla^2h+\eta_1,
\label{er1}
\end{equation}
\begin{equation}
{\partial \phi\over\partial t}+\lambda_3(\nabla h).(\nabla\phi)=\mu\nabla^2\phi
+\eta_2.
\label{er2}
\end{equation}
We choose $
D_1(k)=D_1+D_{\rho}k^{-\rho},\;\; D_2(k)=D_2+D_sk^{-s},$
with $D_1,D_2,D_{\rho}$ and $D_s$ being constants.
We also assume $|\tilde{D}({\bf k})|\sim \tilde{D}k^{-y}$.
We are interested in the long-time, long-wavelength properties of the model;
i.e., we want to calculate the two roughness exponents $\chi_u$ and $\chi_b$
of $\bf u$ and $\bf b$ and the dynamic exponent $z$. 
We employ a standard dynamic renormalisation group procedure. Due to 
Galilean invariance, none of the nonlinearities will renormalise \cite{jkb}.
Thus, if we were to carry out a renormalisation-group
transformation by integrating out a shell of modes
$\Lambda e^{-\ell}<q<\Lambda$, and rescaling ${\bf r} \rightarrow e^{\ell}
{\bf r} , \, {\bf u} \rightarrow e^{\ell \chi_u} {\bf u}, \, {\bf b} 
\rightarrow e^{\ell \chi_b} {\bf b}, \, t \rightarrow e^{\ell z} t$,
the couplings $\lambda_1$ and $\lambda_2$ would be affected
only by the rescaling:
$\lambda_1\rightarrow e^{\ell (\chi_u+z-1)}\lambda_1,
\lambda_2\rightarrow e^{\ell (2\chi_b-\chi_u+z-1)}\lambda_2,$ implying
$\chi_u=\chi_b=1-z.$
These are true independent of the forms of $D_1(k),\,D_2(k)$ and
$\tilde{D}(\bf k)$. Note that the roughness exponents $\chi_1$ and $\chi_2$
of the fields $h$ and $\phi$ are related to those of $\bf u$ and $\bf b$ 
by $\chi_u=\chi_1-1,\; \chi_b=\chi_2-1$.

\begin{figure}
\onefigure[height=3cm]{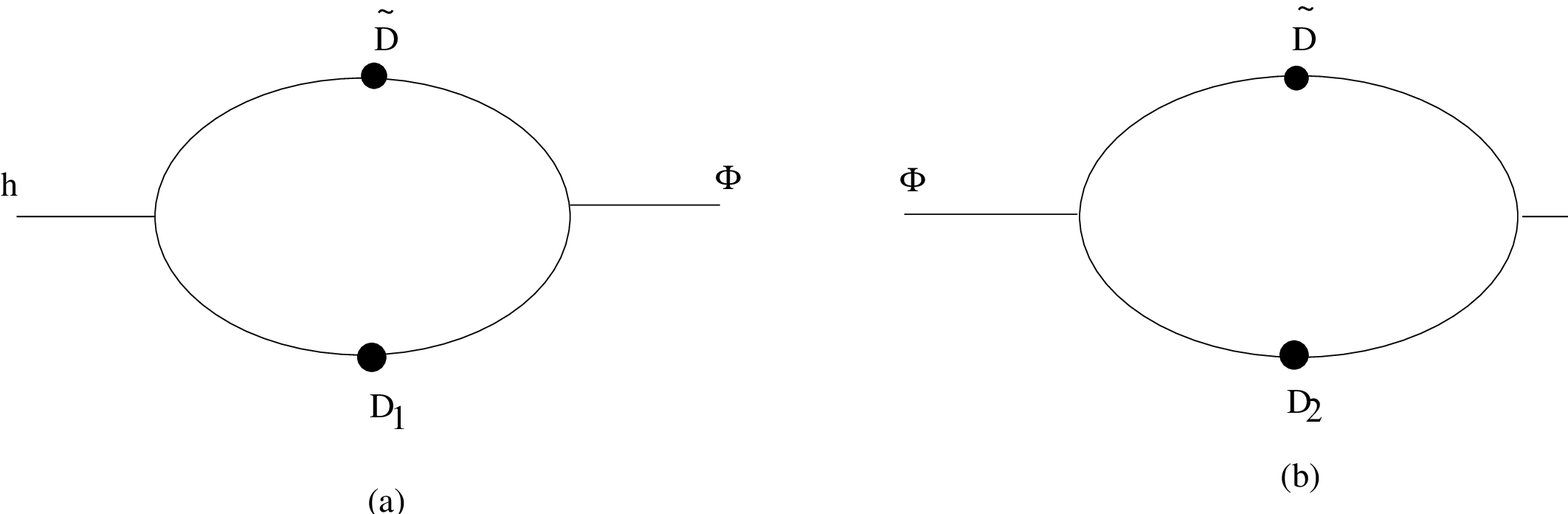}
\caption{One-loop diagrams contributing to the renormalisation of
$\tilde{D}$. Each of them is individually zero (see text).}
\label{dub}
\end{figure}

\begin{figure}
\onefigure[height=3cm]{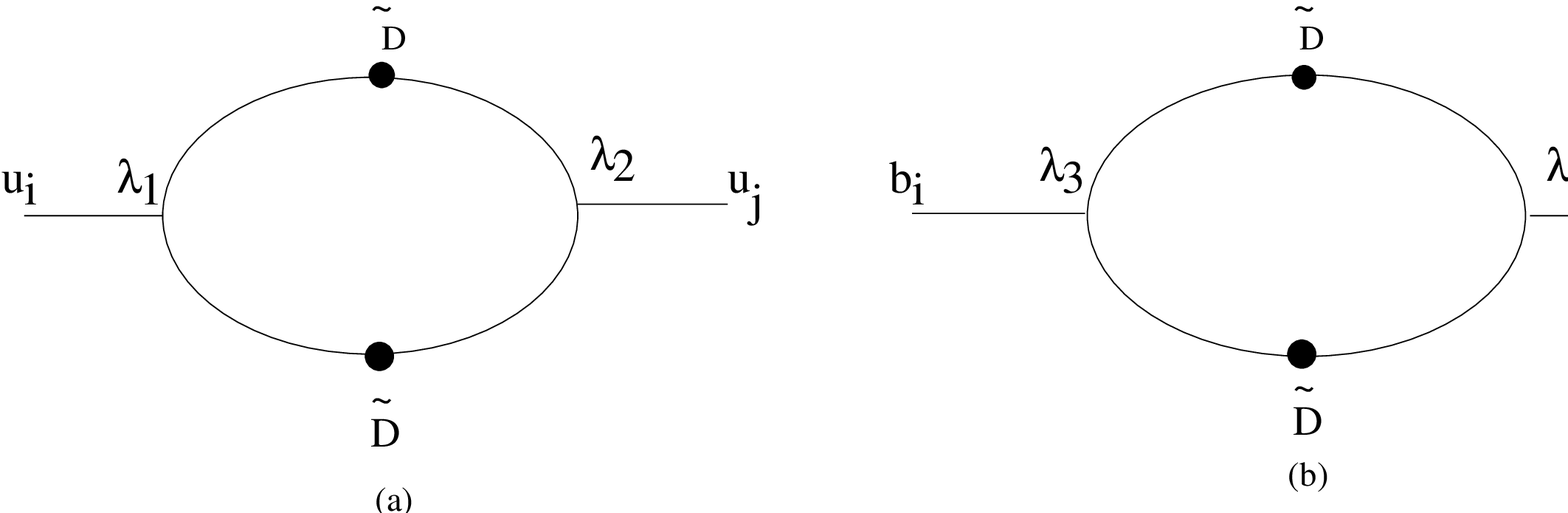}
\caption{A figure showing additional one-loop diagrams renormalising (a) $D_1$
and (b) $D_2$ respectively.}
\label{extra}
\end{figure}

At the beginning let us assume $D_{\rho}=D_s=\tilde{D}=0$,
i.e., the noises are short ranged and there is no crosscorrelation.
We define two dimensionless coupling constants $U\equiv{\lambda^2D_1\over\nu}$
and $V\equiv{\lambda^2D_2\over\nu}$. The nontrivial fixed point is given by
$U=V={2d(d-2)\over 2d-3}$ which is stable at $d=1$ implying a rough 
phase for both $h$ and 
$\phi$ correlations and unstable at $d=2+\epsilon$ indicating a 
smooth-to-rough transition for both $h$ and $\phi$ correlations; this is 
exactly the KPZ universality class and the fixed point is the standard KPZ fixed
point. At the fixed point $\nu_1=\nu_2,\;D_1=D_2,$
giving in $1d$, $z=3/2,\;\chi_u=\chi_b=-1/2$ and in dimensions $d=2+
\epsilon$, $z=2+O(\epsilon)^2,\;\chi_u=\chi_b=-1-O(\epsilon^2)$. These results
imply that at the fixed point the model equations decouple in terms of the
Els\"asser \cite{els} variables $z^{\pm}=u\pm b$ each of which obeys usual 
Burgers
equation with $\langle f^+f^-\rangle=0$ ($f^+,f^-$ are the noises
in the equations for $z^+,z^-$), i.e., noises also decouple in this
new representation. 
Next we make $D_{\rho}$ and $D_s$ nonzero. In $1d$
$\rho,\,s<1/4$ the long-range part of noise
correlations are irrelevant - this is very similar to Ref.\cite{medina} 
(see also Ref.\cite{frey} for a detail discussion). However
when any of $\rho,\,s>\theta$ situation changes drastically. It is usually
taken that the long-range component of the noise does not get renormalised
under the application of RG. It is however, easy to see that if $\rho
\neq s$ this will no longer be true; the less singular one will be 
renormalised to become as singular as the other one. 
At the fixed point we again find $\nu=\mu$. Using Galilean invariance
and nonrenormalisation conditions on the long-range components of the noises
the exponents can be calculated {\em exactly} in dimension $d$
\cite{medina,frey}: We find
\begin{equation}
z=1+{d+1-2\rho\over 3},\;\;\chi_1\,=\chi_2\,={2\rho-d-1\over 3}.
\end{equation}
At this long-range noise fixed point the model again decouples into to
two Burgers equations in terms of the $z^{\pm}$ variables as defined 
above. However the noise correlations do not necessarily decouple as 
in general $D_{\rho}\neq D_s$, i.e., $\langle f^+\,f^-\rangle\neq 0$.
Let us now examine the scaling properties in presence
of a nonzero crosscorrelation, i.e., $\tilde{D}\neq 0$. We first
consider the case when $D_{\rho}=D_s=0$. The procedure of one-loop 
calculations remain same. There are one-loop diagrams (see Fig.\ref{dub}) 
which in principle 
renormalise $\tilde{D}$, however their numerical values are zero, i.e., 
$\tilde{D}$ does not renormalise. 
Each of the diagram has the structure 
\begin{equation}
\int {d\Omega\over 2\pi}{d^dq\over 2\pi}{D_1({\bf q})i\tilde{D}({\bf q})\over
(\Omega^2+\nu q^4)^2}=0
\end{equation}
as the integral is odd in $q$ (since $\tilde{D}({\bf q})$ is a pseudo-scalar 
and odd in $\bf q$).
Due to a non zero $\tilde{D}$ there are now additional contributions to 
both $D_1$ and $D_2$ (see Fig.\ref{extra}), 
with same magnitude but opposite signs. 
clear whether $D_1=D_2$ at the RG fixed point. However, recursion relations
for $\nu$ and $\mu$ remain unchanged. Hence at the RG fixed point $\nu=\mu$.
 We define a new dimensionless coupling constant
$\tilde{U}\equiv {\lambda^2 D_2\over\nu^3}$ in addition to $U$ and $V$ which
we have defined earlier. The flow equations for the couplings are
\begin{equation}
{dU\over dl}= U(2-d)+U^2{2d-3\over 2d}+{V^2\over 4}+{\tilde{U}^2\over 2},
\label{fl1}
\end{equation}
\begin{equation}
{dV\over dl}=V[2-d+{U\over 2}+3U{d-2\over 4d}+3V{d-2\over 4d}]-
{\tilde{U}^2\over 2},
\label{fl2}
\end{equation}
and
\begin{equation}
{d\tilde{U}\over dl}=\tilde{U}[2-d+2y+3U{d-2\over 4d}+3V{d-2\over 4d}].
\label{fl3}
\end{equation}
Let us analyse the case when the crosscorrelations are short ranged, i.e.,
$y=0$. It is easy to see that the nontrivial fixed point solutions of the
Eqs.\ref{fl1},\ref{fl2},\ref{fl3} are the same as the case when there were no 
crosscorrelations: $U=V={2d(d-2)\over 2d-3}$ and $\tilde{U}=0$, 
representing again a rough 
phase at $d=1$ and smooth-to-rough transitions at $d>2$. Interestingly these
fixed points are marginal in $\tilde{U}$ in a linear stability analysis. 
All these suggest that a two loop calculation is necessary to calculate
the full phase diagram. In $1d$ this holds good till $y$ crosses 
the critical value 1/4 as given above.  
As $y$ becomes larger than 1/4, system crosses over to
the long range noise fixed point with the exponents given by the {\em exact}
values (in any $d$)
\begin{equation}
z=1+{d+1-2y\over 3},\;\;\chi_u\,=\,\chi_b={2y-d-1\over 3}.
\end{equation}
Note that at the one-loop level fluctuation corrections to $D_1$ due to the
crosscorrelations (i.e., $\tilde{D}$) is positive, i.e., $\langle uu\rangle$
autocorrelation is enhanced due to $\tilde{D}$, whereas it is negative for
$D_2$, i.e., $\langle bb\rangle$ autocorrelation is reduced from its bare 
value. Since ${\bf b}({\bf r},t)$ is real $\langle {\bf b}_k {\bf b}
_{-k}\rangle$ is positive definite. This suggests that in this model,
for positive definiteness of $\langle {\bf b}_k {\bf b}_{-k}\rangle$ one cannot
have arbitrarily large $\tilde{D}$. This property is also shared by $3d$MHD
equations \cite{abhik}.
If one also introduces long range autocorrelations, i.e., one has nonzero
$D_{\rho},\,D_s$ in the problem then when all of $y,\rho,s$ are less
than $\theta$ the KPZ fixed point is stable. However if any of
$y,\rho,s$ is larger than $\theta$ then the system crosses over to the long 
range noise fixed point like before (for positivity of 
$\langle {\bf b}_k {\bf b}_{-k}\rangle$ we must have $\rho\geq s$ in this 
model. If all or more than one of them
is larger than $\theta$ then the exponents are goverened the largest
of them. Here also an inetersting point is that if $y$ is larger than $\rho$
or $s$ then both $D_{\rho}$ and $D_s$ renormalise such that renormalised
auto correlations scale as $k^{-y}$; however if $y<\rho,s$ then 
crosscorrelation does not renormalise (this is again due to the oddness
of the crosscorrelation). All these suggest that crosscorrelation can 
be a relevant operator in a a field theory for this problem.

Having considered the weak coupling limit for $d>2$ we now examine the
strong coupling phase by using a one-loop selfconsistent scheme.
In particular we investigate the dependence of the ratio $D_2/D_2 (\sim 
E_b/E_v)$ on the crosscorrelations. For simplicity we work with short-ranged 
noises only. We also calculate the upper critical
dimension ($d_c$) of the model in presence of short range crosscorrelations. 
The model is only logarithmically
rough above $d_c$. The value of $d_c$ is still is not well settled 
for the KPZ equation. Some workers suggest it to be $\infty$ \cite{tu}; 
some others suggest a finite $d_c$: L\"{a}ssig and 
Kinzelbach \cite{lassig} mapped KPZ
equation onto a problem of a directed polymer in a random medium and 
showed that $d_c\leq 4$; Bhattacharjee \cite{jkb2} used a mode coupling
approach to show that $d_c=4$. We follow Ref.\cite{jkb2} closely to find 
whether $d_c$ depends upon $\tilde{D}$. 
For simplicity we work with the height variable representation of our model
equations, i.e., with Eqs.\ref{er1} and \ref{er2}.
We assume scaling forms
$G_h=k^{-z}g_h(\omega/k^z);\,C_h=k^{-d-2\chi_1-z}f_h(\omega/k^z);\;
G_{\phi}=k^{-z}g_{\phi}(\omega/k^z);\;C_{\phi}=k^{-d-2\chi_2-z}
f_g(\omega/k^z);\;
\tilde{C}=\tilde{D}({\bf k})k^{-d-\chi_1-\chi_2-z}\tilde{f}(\omega/k^z),$
with $\tilde{D}({\bf k})=-\tilde{D}({\bf -k})$ and $|\tilde{D}({\bf k})|=
\tilde{D}$, a constant, i.e., $\tilde{C}$ is a pseudoscalar in $k$.
Due to Galilean invariance of the nonliearities $\chi_1=\chi_2=
\chi$. From diagramatics, we have seen that one-loop corrections to $\nu$
and $\mu$ are same. Hence we put $\nu=\mu$ without any loss of generality.
This implies $\Sigma_1({\bf k},\omega)=\Sigma_2({\bf k},\omega)=
\Sigma({\bf k},\omega)$.
We assume the forms for the zero-frequency response and correlation functions as
\begin{equation}
\Sigma({\bf k},\omega=0)=\Gamma\,k^z,\,
C_1({\bf k},\omega=0)=D_1k^{-2\chi-d-z},\,
C_2({\bf k},\omega=0)=D_2k^{-2\chi-d-z}.
\end{equation}

We employ a small $\chi$ expansion as used in Ref.\cite{jkb2}.
We calculate self-consistent expressions for the self energies and
correlation functions: Matching at $\omega=0$ \cite{jkb2}
\begin{eqnarray}
{\Gamma^2\over D_1\lambda^2}&=&{1\over 2}\int {d^dp\over (2\pi)^d}
{[{\bf (1-p).p}]^2\over p^{d+2\chi}{\bf (k-p)^{d+2\chi}}}(1+{D_2^2\over
D_1^2}+2{{\tilde D}^2\over D_1^2}).\nonumber \\
{\Gamma^2\over D_2\lambda^2}&=&\int {d^dp\over (2\pi)^d}
{[{\bf (1-p).p}]^2\over p^{d+2\chi}{\bf (k-p)^{d+2\chi}}}
({D_1\over D_2}-{{\tilde D}^2\over D_2^2}).
\end{eqnarray}
Since there is no one-loop diagrammatic correction to $\tilde{C}$
there is no self energy correction for $\tilde{C}$. To the leading
order, from one-loop self energy one obtains 
\begin{equation}
{\Gamma^2\over D_1\lambda^2}={S_d\over (2\pi)^d}{1\over 2d}(1+{D_2\over D_1})
,\;\;\;
{\Gamma^2\over D_2\lambda^2}={S_d\over (2\pi)^d}{1\over 2d}(1+{D_1\over D_2}).
\label{selfen}
\end{equation}
where $S_d$ is the surface of a $d$-dimensional sphere. On the other
hand, one-loop expression for correlation function gives, by extracting the 
high momentum parts ($p>>1$) following Ref.\cite{jkb2}
\begin{equation}
{\Gamma^2\over D_1\lambda^2}={1\over 4}{S_d\over (2\pi)^d}{1\over
p^{d-2+3\chi}}[1+({D_2\over D_1}^2)+2({\tilde{D}\over D})^2],
{\Gamma^2\over D_2\lambda^2}={1\over 4}{S_d\over (2\pi)^d}{1\over
p^{d-2+3\chi}}[{D_1\over D_2}-({{\tilde D}^2\over D_2})^2].
\label{selfcr}
\end{equation}
From Eqs.\ref{selfcr} we find
\begin{equation}
{D_2^2\over D_1^2}+2{D_1\over D_2}\beta +2\beta-1=0,
\label{d1byd2}
\end{equation}
where $\beta\equiv({\tilde{D}\over D_1})^2$. Notice that for 
$\beta=0$ ${D_2\over D_1}=1$. Since one-loop Eq.\ref{d1byd2}
is correct upto $O(\beta)$, we look for solution for $D_2\over D_1$ of the
form ${D_2\over D_1}=1+a\beta$, such that for $\beta=0$ we recover $D_1=D_2$.
We obtain $a=-2$,i.e., $D_2/D_1=1-2\beta$. So within this
approximate calculation $\beta$ cannot exceed 1/2 (i.e., $\tilde{D}\leq D_1/2$
). This immediately gives $\chi_1=\chi_2=1/2+O(\beta^2)$ in $d=1$ in 
agreement with the one loop DRG results: $\chi_1=\chi_2=1/2$ which are 
correct to $O(\beta)$. Following Ref.\cite{jkb2} we obtain $d_c=4+O(\beta)^2$. 
These results also
suggest the need of a two-loop calculation. 

We now discuss the effects of the antisymmetric part of the crosscorrelation
(we now work with Eqs.\ref{jkb1u},\ref{jkb1b} as, in the presence of 
antisymmetric crossocrrelations, Eqs.\ref{jkb1u},\ref{jkb1b} do not reduce to
Eqs.\ref{er1} and \ref{er2}. However, the procedure is identical.).
We have
\begin{equation}
D_{ij}^a({\bf k})=-D_{ij}^a({\bf -k})=D_{ji}^a({\bf -k})=-[D_{ij}^a({\bf k})]^*.
\end{equation}
We choose $D_{ij}^a({\bf k})D_{ji}^a({\bf -k})=\hat{D}^2k^4$ and $\tilde{D}=0$.
When the crosscorrelation is antisymmetric (i.e., $\tilde{D}=0,\,\hat{D}
\neq 0$) fluctuation corrections to both $D_1$ and $D_2$ are positive and
same value: The flow equations for dimensionless coupling constants
$U,\,V$ and $\hat{U}\equiv {\lambda^2\hat{D}\over\nu^3}$ are
\begin{equation}
{dU\over dl}= U(2-d)+U^2{2d-3\over 2d}+{V^2\over 4}+{\hat{U}^2\over 2},
\label{fl11}
\end{equation}
\begin{equation}
{dV\over dl}=V[2-d+{U\over 2}+3U{d-2\over 4d}+3V{d-2\over 4d}]+
{\hat{U}^2\over 2},
\label{fl22}
\end{equation}
and
\begin{equation}
{d\hat{U}\over dl}=\hat{U}[2-d+2y+3U{d-2\over 4d}+3V{d-2\over 4d}].
\label{fl33}
\end{equation}
Since the fluctuation corrections to $D_1$ and $D_2$ due to $\hat{D}$ are
positive, there is no restriction on the strength of $\hat{D}$, unlike
the case with symmetric crosscorrelations.
The exponents at the weak coupling fixed point for $d>2$ remain unaffected 
by $\hat{U}$ at the one loop level (at $1d,\,\hat{U}\equiv 0$). It is easy
to find out $d_c$ when $\hat{D}\neq 0$ following Ref.\cite{jkb2}. We find 
$D_1=D_2$ at the strong coupling fixed point and $d_c=4+2\beta$, keeping
terms only at the linear order in $\beta$. Thus in presence of an additional
noise in the form of the crosscorrelations, $d_c$ and hence the roughness
(at $d_c$ the roughness exponent drops to zero from a positive value)
increases.

We have seen that when $\tilde{D}=0$ (i.e., there are no crosscorrelations)
$D_1=D_2$ (or $U=V$, by assuming $\nu=\mu$) at the RG fixed point. In fact,
$U=V$ is the FDT line in the ($U-V$) plane. If we start with $D_1=D_2=D$, this
remains invariant under the RG transformations. Along this line the model
equations decouple into two independent Burgers equations in terms of
$z^+,z^-$. In $1d$ along the line $U=V$ $\nu_I/\nu=D_I/D$, telling immediately
that the Fluctuation-Dissipation Theorem (FDT) \cite{fns} holds. Thus we get
the steady state probability distribution $P[u,b]\sim \exp[{\nu\over D}
\int_x \{(\partial_x u)^2+(\partial_x b)^2\}]$. We 
immediately obtain $\chi_u=\chi_b=-1/2$ {\em exactly} and hence $z=3/2$.
However in $1d$ when $\tilde{D}\neq 0$  the exponents do not depend upon 
$\tilde{D}$ at the one-loop level, FDT does not hold any longer,
simply because there are
additional fluctuation corrections to $D_1$ and $D_2$ due to $\tilde{D}$:
\begin{equation}
\nu_I=\nu(1+{\lambda^2D_1\over 4\nu^3}+{\lambda^2D_2\over 4\mu_o^3}),\,
\mu_I=\mu(1+{\lambda^2D_1\over 4\nu^3}+{\lambda^2D_2\over 4\mu_o^3}),
\label{fdt1}
\end{equation}
\begin{equation}
D_{1I}=D_1[1+{\lambda^2D_1\over 4\nu^3}+
{\lambda^2(D_2)^2\over 4D_1\mu^3}+2{\lambda^2 D_o\over 4\nu^3}({\tilde{D}\over
D_1})^2 ],\,
D_{2I}=D_2[1+2{\lambda^2D_1\over 4\nu^3}-
2{\lambda^2 D_2\over 4\nu^3}({\tilde{D}\over D_1})^2({D_2\over D_1})^2].
\label{fdt3}
\end{equation}
Obviously $\nu_I/D_{1I}\neq \nu_o/D_1$ and $D_2/D_1=1-2\beta\neq 0$, 
consequently the FDT does not hold good. Thus $D_1=D_2$ is maintained
under RG transformations. Hence the exact values of $z,\chi_1,\chi_2$ are not
known, unlike the case with $\tilde{D}=0$.
Recall that when $\tilde{D}=0$, $\nu_I/D_{1I}=\nu/D_1$, which is nothing but a
statement of the FDT \cite{fns}. Note that the dimensionless ratio $({\tilde{D}
\over D_1})^2$ determines the amount of deviation from the FDT.

What are  the implications of the above mentioned results on
fully developed MHD turbulence. In 
MHD turbulence in general crosshelicity is nonzero and found to scale
as $\sim k^{-5/3}$ (within error bars) \cite{thesis}. This makes it 
imperative to have nonzero crosscorrelations of noises in stochastically driven
MHD models (only the symmetric part contributes to the crosshelicity, however, 
the existence of an antisymmetric part cannot be ruled out on the basis of any
general principle). $k^{-5/3}$-spectrum
requires noise correlation to scale as $k^{-3}$ in $3d$ (for both auto
and crosscorrelations). If all bare correlations scale in the same way then
one loop corrections to the autocorrelations at zero external frequency
scale as the bare ones in the low momentum limit; however one-loop
correction to the crosscorrelation is identically zero. 
This means the energy spectrum is at least as singular as (or
more than) the crosshelicity spectrum. It is interesting that similar behaviour
holds good also for the stochastically driven $3d$MHD equations \cite{abhik}.
We have also seen that in our calculations, the ratio $D_2/D_1$ depends upon
$\tilde{D},\hat{D}$, the strength of crosscorrelations. 
Now $E_b/E_v\sim D_2/D_1\sim 1-2\beta$.
Thus in this model, by varying $\tilde{D}$ and $\hat{D}$ we can 
achieve different values of
$E_b/E_v$, ranging from 0 to 1. Similar dependence of $E_b/E_v$ on 
$\tilde{D}$ holds for $3d$MHD
equations also \cite{abhik}. Of course, to model MHD turbulence one must
work with noise correlations that are singular in the $k\Rightarrow 0$ limit.
For such noises, scaling (i.e., the $k$-dependence) of the noise 
correlations do not change under renormalisation, but the amplitudes change
in such a way that $D_1$ is enhanced and $D_2$ is suppressed in presence of a
crosscorrelation, affecting the ratio $E_b/E_v$.
Thus, in conclusion, we have discussed the role of a crosscorrelation
of noises on the scaling properties of the reduced
model for MHD. In particular we have shown that sufficiently singular
crosscorrelation can be a relevant perturbation on the model. The upper critical
dimension of the model in presence of short range noise only does
depend upon the strength of the crosscorrelations and, interestingly,
the dependence is different for the symmetric and the antisymmetric parts. 
In the presence of antisymmetric crosscorrelations, the reduced model
does not reduce to that of Ert\c{a}s and Kardar. However one cannot rule out
its presence in the context of MHD on any symmetry grounds.  
It is important to examine the effects of crosscorrelations (thus finite
crosshelicity) on the values of the universal numbers in fully developed
MHD turbulence, e.g., Kolmogorov constants, intermittency exponents 
\cite{abhik}. It would be very interesting to check our results in numerical
simulations of our model and/or $3d$MHD equations.


\end{document}